\documentclass[final,authoryear,3p,times]{elsarticle}
\usepackage{epsfig}

\usepackage{amssymb}
\usepackage{journalnames} % This is the name of the .sty
\usepackage{txfonts}
\usepackage{color}

\journal{New Astronomy}

\def\astrobj#1{#1}

\newcommand{\HS}{\astrobj{HS~1603+3820}}

\newcommand{\kms}{\mbox{\rm km\thinspace s$^{-1}$}}
\newcommand{\cmd}{\mbox{\rm cm$^{-2}$}}
\newcommand{\cmt}{\mbox{\rm cm$^{-3}$}}

\newcommand{\NH}{$N_{\rm H}$}
\newcommand{\NHt}{$N_{\rm H}^{tot}$}

\newcommand{\HI}{H{\sc i}}
\newcommand{\HeI}{He{\sc i}}
\newcommand{\CII}{C{\sc ii}}
\newcommand{\CIV}{C{\sc iv}}
\newcommand{\SiII}{Si{\sc ii}}

\newcommand{\SiXIV}{Si{\sc xiv}}
\newcommand{\NV}{N{\sc v}}

\newcommand{\OVI}{O{\sc vi}}
\newcommand{\OVIII}{O{\sc viii}}
\newcommand{\AlII}{Al{\sc ii}}

\newcommand{\MgII}{Mg{\sc ii}}
\newcommand{\MgXII}{Mg{\sc xii}}
\newcommand{\FeII}{Fe{\sc ii}}

\newcommand{\FeXXV}{Fe{\sc xxv}}
\newcommand{\NC}{$N_{\CIV\ }$}
\newcommand{\NHI}{$N_{\HI\ }$}
\newcommand{\NCH}{$N_{\CIV\ }/N_{\HI\ }$}

\newcommand{\ttil}{\mbox{\char'176}}

\begin{document}

\begin{frontmatter}

\title{Absorption features in the quasar \astrobj{HS 1603+3820} II.   
     Distance to the absorber from photoionisation modelling}

\author[camk]{A. R{\'o}\.za{\'n}ska \corref{cor1}}
\ead{agata@camk.edu.pl}

\author[uwb]{M. Niko\l ajuk}
%\ead{mrk@alpha.uwb.edu.pl}

\author[camk]{B. Czerny}
%\ead{bcz@camk.edu.pl}

\author[eso]{A. Dobrzycki}
\author[camk]{K. Hryniewicz}
\author[bech]{J. Bechtold}
\author[ebe]{H. Ebeling}

\cortext[cor1]{Principal corresponding author}

\address[camk]{N. Copernicus Astronomical Center, Bartycka 18, 00-716
  Warsaw, Poland} 
\address[uwb]{Faculty of Physics, University of Bialystok, Lipowa 41,
  15-424 Bialystok, Poland}
\address[eso]{European Southern Observatory,
  Karl-Schwarzschild-Strasse 2, D-85748 Garching bei M\" unchen,
  Germany}
\address[bech]{Steward Observatory, University of Arizona, Tucson, AZ
  85721, USA}
\address[ebe]{Institute for Astronomy, University of Hawai`i, 2680
  Woodlawn Drive, Honolulu, HI 96822, USA }

%------------------------------------------------------------------------------------------------------
\begin{abstract}
  We present photoionisation modelling of the intrinsic absorber in the
  bright quasar \HS. We construct broad-band spectral energy
  distribution using optical/UV/X-ray observations from different
  instruments as an input to photoionisation calculations.  Spectra
  from Keck telescope show extremely high ratio of \CIV\ to \HI\, for
  the first absorber in system A, named A1. This value, together with
  high column density of \CIV\ ion, puts strong constraints on
  photoionisation model.  We use two photoionisation codes to derive
  hydrogen number density at the cloud illuminated surface.
  Estimating bolometric luminosity of \HS, from typical formula for
  quasars, we calculate the distance to A1.  Either for constant
  density cloud (modelled by {\sc cloudy}), or stratified cloud
  (modelled by {\sc titan}) we were able to find a single
  photoionisation solution, assuming solar abundances, which explains
  both the ionic column density of \CIV, and high \CIV\ to \HI\ ratio.
  The derived location is as close as 0.1 pc, and situates an absorber
  even closer to the nucleus than the possible location of  the Broad
  Line Region in this
  object. The upper limit for the distance is sensitive to the adopted
  covering factor and the carbon abundance.  Photoionisation modelling
  always prefers dense clouds with  the number density
  $n_0= 10^{10}$, $10^{12}$~\cmt, to
  explain observational constrains of \HS\ intrinsic absorption. This
  number density is of the same order as the number density in the
  disk atmosphere at the implied distance of the A1.  Our results thus
  show that the disk wind escaping from outermost accretion disk
  atmosphere can build up dense absorber in quasars.
\end{abstract}


\end{frontmatter}

% \linenumbers

%%%%%%%%%%%%%%%%%%%%%%%%%%%%%%%%%%%%%%%%%%%%%%
\section{Introduction}
\label{sec:intro}

High resolution optical/UV spectra show that over 20\% of quasars
exhibit broad absorption lines (BALs) from ionised species of heavy
elements.  In some objects lines are from high-ionisation species, as
\CIV, \NV, \OVI\ forming subclass of HiBALs, and in other cases
absorption features originate from low-ionisation species as \MgII,
\AlII, \SiII, \FeII\ forming LoBALs subclass.

In all BALs, lines can exhibit very complex blushifted profiles,
indicating the presence of several absorbing systems moving toward an
observer with different velocities, for example in QSO 2359-1241
\citep{arav2001}. Narrow line components indicate usually velocities
of the order of a few hundred \kms, while velocities of some broad
lines can reach even 50,000~\kms\ \citep{weymann95}.

Similar absorption in UV is observed in Seyfert galaxies
\citep{crenshaw1999}.  Moreover, in 50\% of those objects, the matter
at higher ionisation state, so called ``warm absorber'', is detected
in X-ray band, for example in NGC 3783 \citep{kaspi2001}. Observed
spectral features from \OVIII, \MgXII, \SiXIV, and even \FeXXV\
indicate temperature of an absorbing gas to be of the order of
$10^6-10^7$~K. Typical ``warm'' X-ray lines are narrower than those of
BALs, and blushifted by velocities of the order of few hundreds \kms\
up to $10^3$~\kms\ \citep[for review, see][]{blustin2005}.

Many questions arise about the origin and physical conditions of those
outflows.  The first obvious is: does the UV and X-ray absorption
occur in the same wind?  There are two examples showing that the line
detected in X-rays can be fitted exactly by the same velocity profile
as the line detected in UV for two active galaxies NGC 5548 and NGC
3783 \citep{kaastra2002,gabel2003}.  In general, to answer this
question we have to collect multiband optical/UV/X-rays observations
of quasars with resolution high enough to compare line profiles from
different energy bands.  However, presently working satellites as {\it
  Chandra} and {\it XMM-Newton} are not able to detect X-ray
absorption in distant quasars, and very often X-ray spectrum can be
modelled only as a power-law, like in case of Chandra detection of
\HS\ \citep[][hereafter Paper I]{dobrzycki2007}.

Physical conditions in UV or X-ray absorbing clouds can be studied by
photoionisation modelling. For that purpose, a broad-band spectral
energy distribution (SED) of a considered source is needed.
Nevertheless, many observations do not cover EUV band.  For this
reason in several recent papers on UV outflows, authors have used
standard \citet{mathews1987} composite quasar spectrum for
photoionisation modelling.  It is widely known, that the SED of
incident photons determines ionisation and thermal structure of the
absorbing plasma \citep{rozanska2008,cha2009}.  In this paper, taking
into account multi-wavelength observations of \HS\ from {\it MMT},
{\it Keck} and {\it Chandra} telescopes we have constructed the SED of the
quasar spectrum. This SED is used as an input for our advanced
photoionisation modelling of the system using codes {\sc cloudy} and
{\sc titan}.

Photoionisation modelling provides us with a possibility to determine
the distance of the cloud from the nucleus, only if the SED is
dominated by soft disk component, as shown in \citet{rozanska2008}.
Photoionised models depend predominantly on the ionisation parameter,
i.e. a combination of the surface number density and a distance from
the source.  Models illuminated by only hard X-ray power-law are
degenerated for the gas number density up to $10^{11}$~\cmt.  In
  thic case, we are not able to differentiate between matter located
closer to the nucleus with higher density, and more diluted gas being
farther away.  The transmitted spectra through such two clouds look
the same, since their ionisation and temperature structure are
identical \citep{rozanska2008}.  In case of \HS,  where
the SED is dominated by soft UV emission over X-rays, this degeneracy
is broken due to the fact that the Comptonization becomes not important
comparing to the free-free processes \citep{rozanska2008}.  Therefore, the
hydrogen number density at the cloud surface can be estimated from the
line ratios, and this density together with quasar luminosity, and
ionisation parameter gives us a distance of an absorber from the
source of continuum.
    
Our method is complementary to other two methods used for distance
derivation, based on different ways of determining the density number
at an absorbing cloud surface. In the first one, it can be done from
variability data, assuming that observed spectral changes are due to
variation of the ionisation state \citep{krolik1995}. This method
allowed us to estimate the number density of the warm absorber in
Seyfert 1 galaxy NGC 4055 from ROSAT data to be of the order of
$n=10^8$~\cmt, implying the location of the gas at about $R=0.0024$~pc
\citep{nicastro1999}.  Nevertheless, in case of NGC 3783 two groups
using this method got opposite results.  \citet{netzer2003} have found
clouds at three different ionisation levels located on 3.2, 0.63, and
0.18~pc from the nucleus, while \citet{krongold2005} found two clouds
on 0.0029, and 0.0004-0.0008~pc, using the same 900~ks Chandra
observations.  Outflowing velocities of that absorbers were of the
order of 750~\kms.

The second method uses high quality UV spectra of LoBALs, since number
density can be estimated by comparing the ionic column densities of
excited metastable states to their ground states of the most popular
ions \HeI, \CII, \SiII, and \FeII\
\citep{korista2008,moe2009,bautista2010}.  In case of three quasars
presented in above papers, estimated number density is always of the
order of $10^3 - 10^{4.4}$~\cmt, indicating the gas location at about
3-6~kpc. The velocities of those absorbers are in the range of
1400-7500~\kms.

In this paper we perform photoionisation modelling of an intrinsic
absorption in quasar \HS\, with the aim to determine physical
properties of an absorbing gas. This HiBAL type object is interesting
since the ratio of \CIV\ to \HI\ was found to be extremely high -
above 20 (Paper I).  The most interesting absorbing systems are
outflowing with velocities reaching 3000~\kms.  By comparison of model
parameters to those indicated from observations of \HS, we are able
to find a single photoionisation solution, which can explain observed
ionic column densities of the first cloud in system A, reported in
Paper I and marked as A1 here.

We use photoionisation modelling to determine uniquely both the
hydrogen number density at the illuminated surface of the cloud, and
the ionisation parameter.  Those two quantities combined with the
source luminosity allow us to compute the distance from the central
nucleus to the absorbing system.  We present such calculations in this
paper and compare results with previous estimations done by
\citet{misawa2005,misawa2007}.

This paper is organised as follows: in Sec.~\ref{sec:src} we present
the source, and in Sec.~\ref{sec:obs} we review its optical/UV/X-ray
observations.  Photoionisation models are presented in
Section~\ref{sec:models}.  The location of an absorber and connection
to an accretion disk atmosphere is discussed in
Sec.~\ref{sec:distance} and Sec.~\ref{sec:blr} respectively.  We
summarise our results in Sec.~\ref{sec:summary}.

%___________________________
\section{The source}
\label{sec:src}

Recently observed high redshift quasar \HS\ with $z_{\rm em} = 2.54$
indicates extreme richness of absorption lines in its spectra
\citep[][Paper I]{dobrzycki1999,misawa2003, misawa2005}.

The source was first discovered in the Hamburg/CfA Bright Quasar
Survey \citep{hagen1995,dobrzycki1996}. With the brightness $B=15.9$,
it is among the top few brightest quasars known at such redshift.
However, the most striking feature of \HS\ is its absorption spectrum,
which indicates the presence of $\sim 50$ individual absorption
systems, with 30+ of them having velocities higher than 10000~\kms,
when corrected for quasar redshift based on emission lines.  Many -
perhaps all - of these clouds are physically associated with the
quasar, particularly because some lines show variability in the
timescales of a few years \citep{misawa2007}.

High resolution observations from Subaru telescope with the High
Dispersion Spectrograph, reported by \citet{misawa2003,misawa2005}
enabled to resolve many of those systems into a number of narrow
components. The authors grouped the absorbers into several systems,
designated with the letters from A to H.

All of those systems, and two previously unknown, were indicated in
the observations taken by MMT and Keck telescope and reported in Paper
I.  Furthermore, the authors have noticed that some clouds in system A
have a high ratio of \CIV\ to \HI\ column densities, reaching $\sim
20$. System A, and other systems spatially close to the quasar can be
used as a probe for the intrinsic emission of the quasar itself, since
the conditions in the systems are undoubtedly heavily influenced by
the quasar flux \citep[e.g.][and references
therein]{crenshaw2003,gabel2005,scott2005}.

Additionally, in Paper I, X-ray {\it Chandra} observations of \HS\
were reported, allowing to calculate the relative optical-to-X-ray
slope, $\alpha_{ox}=1.49$, which is typical for a quasar at this
redshift \citep{bechtold2003}.

%_____________________________________________________________
\section{Observational constrains important for 
photoionisation modelling}
\label{sec:obs}

To explain the properties of an intrinsic absorption in \HS\, we need
to construct illuminated continuum which affects absorbing gas.  The
observational data used in this paper are described in detail in Paper
I.  In subsections below, we shortly remind the reader values of
fitted parameters for those observations, and summarise what is
important for our modelling.

\subsection{Continuum shape and normalisation}
\label{sec:obsxray}

%%%%%%%%%%%%%%%%%%%%%%%%%continuum figure 

\begin{figure}
\epsfxsize=8.8cm \epsfbox[80 350 625 700]{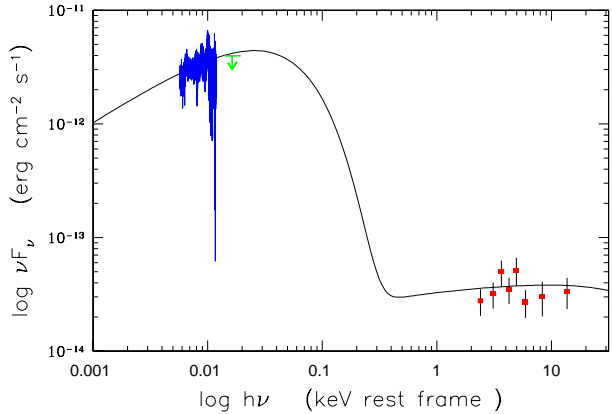}
\caption{Normalised multi-band spectrum of \HS\ from MMT and Keck
  (blue points), IUE upper limit (green arrow), and Chandra (red
  squares) in the rest frame of quasar.  Solid line represents the
  best shape of modelled spectrum used in calculations of photoionised
  intrinsic absorber presented in Sec~\ref{sec:models}.}
\label{fig:continuum}
\end{figure}

The optical/UV data spectrophotometry was taken from
\citet{scott2000}, and it was obtained on July 4, 1995 with Stewart
Observatory Bok Telescope.  Our continuum fit results with power-law
photon index $\Gamma_{UV} = 1.36 \pm 0.1$. Observed optical flux at
$\lambda_{rest}=1450$~\AA\ ($E_{rest}=8.55 \times 10^{-3}$ keV) is $ E
F_{E}(B)= 3.20 \times 10^{-12}$~erg~s$^{-1}$~\cmd\ as reported by
\citet{scott2000} (see their Tab. 3).  In Fig.~\ref{fig:continuum},
blue points represent overall normalised flux in optical/UV domain.

The X-ray data were taken with the {\em Chandra X-ray Observatory} on
2002 November 29. We performed the fit assuming the intrinsic quasar
spectrum to be a power law, with fixed Galactic absorption towards
\HS\ of \NH~$=1.3\times10^{21}$~\cmd.  The best fit gave the photon
index of $\Gamma_{X}=1.91 \pm0.20$ and normalisation at the rest frame
for 1~keV of $A=(2.43\pm 0.79)\times 10^{-4}$
photons~\cmd~s$^{-1}$keV$^{-1}$.  The X-ray part of observed and
modelled spectrum is presented in the Fig.~\ref{fig:continuum}, red
squares.

\HS\ has been pointed at with IUE/LWP on 1995-08-19, but not detected.
The observation gave the upper limit of $1.5 \times
10^{-15}$~erg~s$^{-1}$~\cmd~\AA$^{-1}$ in the 2200-3100~\AA\
(observed) wavelength range. This limit is marked in green on
Fig.~\ref{fig:continuum}. We note that this wavelength range covers
the inferred position of Lyman-limit absorption from associated
absorbers. It is therefore possible that this limit does not reflect
low QSO's intrinsic luminosity in this range, although the individual
absorbers do not appear to have sufficient column densities.

We adopt those values for all photoionisation models in
Sec.~\ref{sec:models}, and for an estimation of bolometric luminosity
in Sec.~\ref {sec:distance}.  We present the overall continuum in
Fig.~\ref{fig:continuum}.  Note that different parts of the spectrum
are taken at different epochs therefore SED constructed by us can
contain systematic errors.  Nevertheless, for the time being, there is
no other way to find continuum for this quasar.
%________________________________________________________________
\subsection{Optical/UV spectroscopy and parameters from spectral lines
  fitting}

\begin{table*}
\begin{center}
  \caption{Ratios of ionic column densities of \CIV\ and \HI\,
    observed in intrinsic absorption system A1, obtained for various
    assumptions about the covering factor of \HI\ cloud and fitting
    methods for system A1 ($z_{abs} =2.41941 \pm 0.00002 $).  The name
    of the system and the No. Line ID are taken after our Paper I.}
\label{tab:ratio}
\begin{tabular}{lcccccrr} % No vertical elements in tables.
\hline \hline 
Sys. No.& log(\NC\ ) & $c_f$ (\NC\ )   & b(\NC\ ) & log(\HI\ ) &  $c_f$ (\HI\ ) & b(\HI\ ) & \NCH\  \\
\hline
 A1$^g$  & $14.97 \pm 0.04$ & 0.36 & $72\pm 3$ & $13.65 \pm 0.03$ & 0.61 & $62 \pm 10 $ & $20.37 \pm 2.70$ \\
 A1      & $14.97 \pm 0.04$ & 0.36 & $72\pm 3$ & $13.87 \pm 0.05$ & 1.00 & $145 \pm 12 $ & $12.4 \pm 2.10$ \\
 A1      & $14.97 \pm 0.04$ & 0.36 & $72\pm 3$ & $14.11 \pm 0.05$ & 0.61 & $132 \pm 10 $ & $ 7.1 \pm 1.20$ \\
 A1      & $14.97 \pm 0.04$ & 0.36 & $72\pm 3$ & $14.34 \pm 0.05$ & 0.36 & $75 \pm 9  $ & $ 4.2 \pm 0.70 $ \\
\hline \hline
\multicolumn{3}{c}{ $^g$ global fit for all lines of system A from Paper I}
\end{tabular}
\end{center}
\end{table*}

\HS\ has been observed with two instruments. On 2001 August 19 we used
the Echelle Spectrograph and Imager \citep{sheinis2002} on the Keck~II
telescope, with the combined exposure of 4800~sec.  The 0.5~arcsec
slit was used, yielding the resolution of 45-50~\kms.

\begin{figure}
\epsfxsize=8.8cm \epsfbox[90 350 625 700]{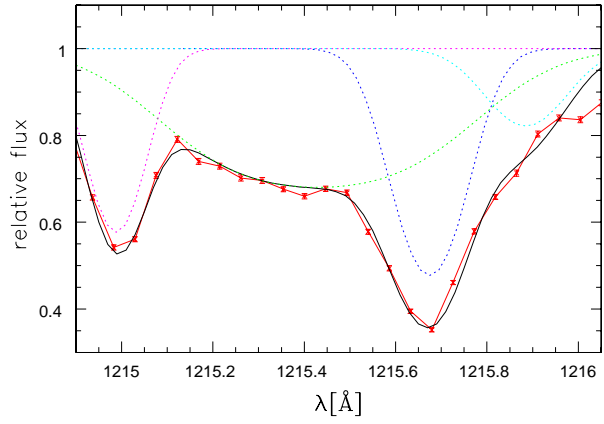}
\caption{The fit of the Ly$\alpha$ region for A1 component taking
  covering factor equal 0.36, as for \CIV\ .  Broad green dotted line
  represents an \HI\ line of intrinsic absorber, other magenta and
  blue components represent Ly$\alpha$ forest, red dots and line are
  observed Keck data, and black continuous line is the final model.}
\label{fig:c36}
\end{figure}

Full procedure of spectral line fitting of optical/UV \HS\ spectrum is
presented in Paper I.  The line profile fits were performed using
VPGUESS/VPFIT
programmes\footnote{http://www.eso.org/{\ttil}jliske/vpguess/,
  http://www.ast.cam.ac.uk/{\ttil}rfc/vpfit.html}.  We were able to
fit simultaneously several lines from different ions, but
characterised by the same redshift.

In several clouds belonging to an intrinsic absorption systems the
column density of \CIV\ ion is bigger than the column density of
ionised hydrogen \HI\ . The cloud A1 was an extreme example, with the
ratio of \NCH\ equal $20 \pm 3$. In the present paper we concentrate
on modelling this component in detail.

\begin{figure}
\epsfxsize=8.8cm \epsfbox[90 350 625 700]{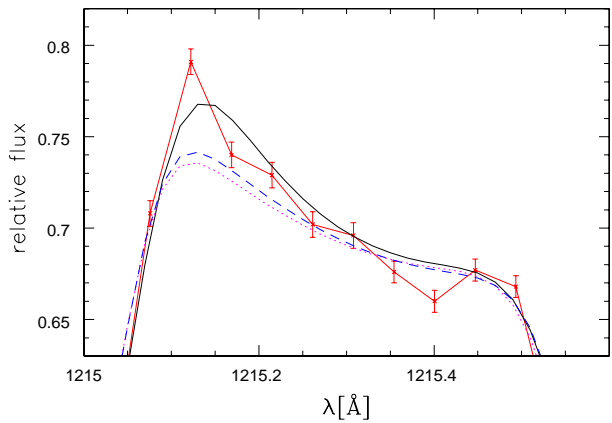}
\caption{Enlarged part of Fig.~\ref{fig:c36} where the Ly$\alpha$
  region of the A1 component is fitted with three different covering
  factors for \HI\ absorber. Solid-black line is for covering factor
  equal 0.36, dashed-blue - 0.61, and dotted-red - 1, i.e. complete
  covering.  Again data from Keck are marked by red points and line. }
\label{fig:set}
\end{figure}

In Paper I, the covering factor for the hydrogen and for carbon
fitting were found to be different (0.36 and 0.61, correspondingly)
which could argue against a single well defined cloud responsible for
A1 absorption.  However, the kinematic width of the lines were
similar.  Therefore, we reconsidered the spectroscopic data by
performing fits specifically for the component A1, instead of assuming
the same covering factor for all clouds in component A.  The old
result and new results are given in Tab.~\ref{tab:ratio}.

In Fig.~\ref{fig:c36} we show that the covering factor implied for CIV
component can provide good fit to the \HI\ component.  Thus both
covering factor and the kinematic width are the same for the two
elements, and the absorption is likely coming from a single cloud. The
fit is actually somewhat better for low covering factor although the
results are not firm since we see well only a fraction of the line due
to overlapping narrower components from Ly$\alpha$ forest (see
Fig.~\ref{fig:set}).

Although the errors are huge, the values of ratio are always much
higher than unity, which indicates either overabundance of carbon, or
peculiar photoionisation model.  For farther analyses we refer to the
value of that ratio equal 20, as found in Paper 1.
%_________________________________________
\section{Photoionisation models}
\label{sec:models}

\begin{figure}
\epsfxsize=8.8cm \epsfbox[50 50 380 820]{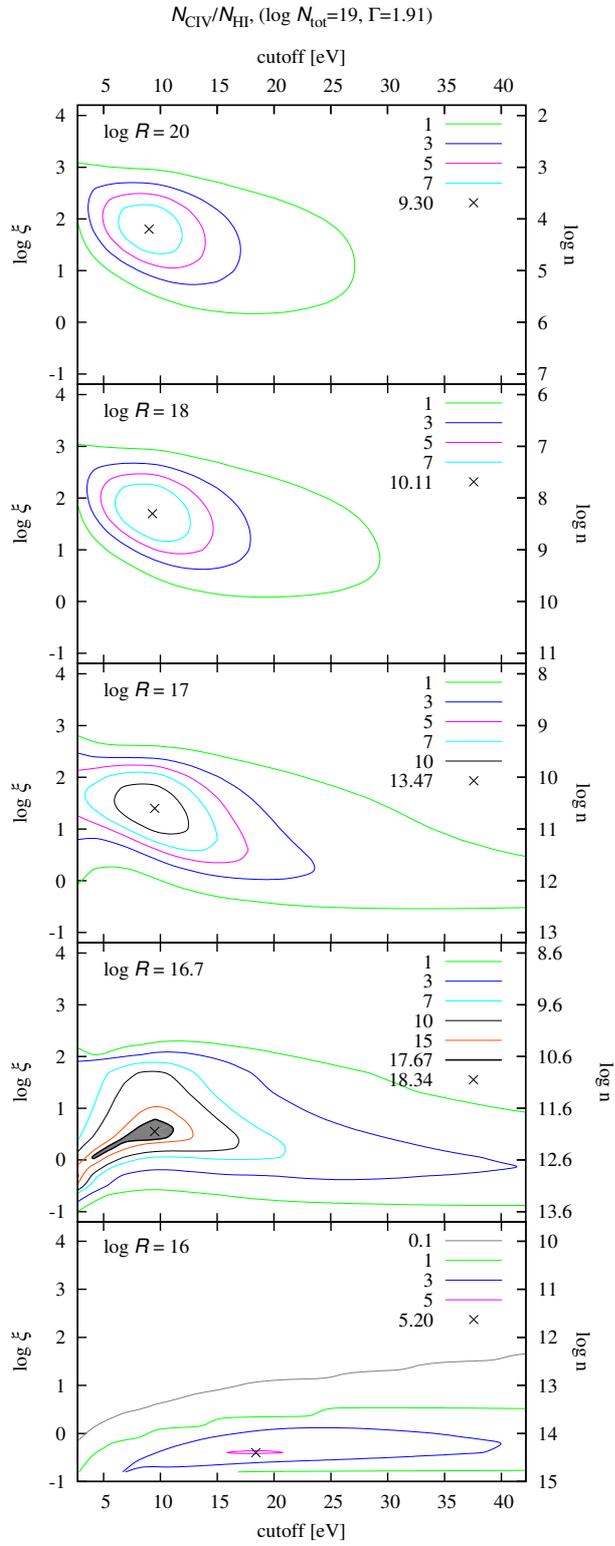}
\caption{Contour plots for the ratio of \CIV\ to \HI\ column densities
  from {\sc cloudy} for clouds with total column density
  log(\NHt~)~$=19$~[\cmd].}
\label{fig:col19}
\end{figure}

\begin{figure}
\epsfxsize=8.8cm \epsfbox[50 50 380 820]{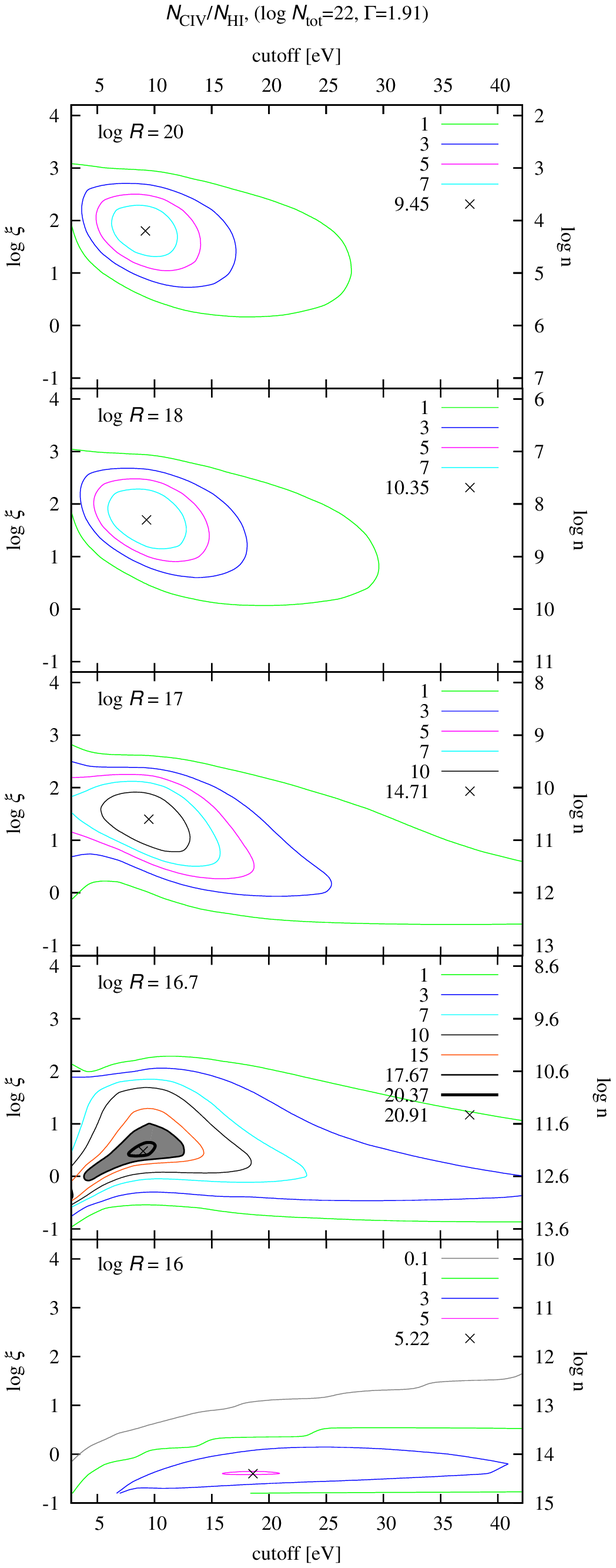}
\caption{Contour plots for the ratio of \CIV\ to \HI\ column densities
  from {\sc cloudy} for clouds with total column density
  log(\NHt~)~$=22$~[\cmd].}
\label{fig:col22}
\end{figure}

In order to explain high ratio of \NC to the \NHI in the absorbing
medium of the \HS\ we perform photoionisation calculations using two
codes: {\sc cloudy} \citep{ferland1998} version 08 and {\sc titan}
\citep{dumont2000}.

We model the broad band spectrum of \HS\ shown in
Fig.~\ref{fig:continuum} using two power law components, both with an
exponential cut-off at low and high energies:
\begin{equation}
 F_{E} = E^{-\alpha}exp(-E/E_{max})exp(-E_{min}/E),
\label{eq:flux}
\end{equation}
where spectral index $\alpha=\Gamma -1$, as usual.  The first power
law represents the optical/UV spectrum: $\alpha_{UV} = 0.36$,
$E_{min}^{UV}=1.36 \times 10^{-4}$~keV, with $E_{max}^{UV}$ being
first free parameter of the model, and the second power law represents
the X-ray emission ($\alpha_X = 0.91$, $E_{min}^{X}=E_{max}^{UV}$,
$E_{max}^{X}=100$~keV).

The important parameter is UV power-law high energy cut-off, since it
regulates amount of soft photons affecting the absorber.  For the
solid line in the Fig.~\ref{fig:continuum},  which represents the
  best shape of modelled spectrum, the value
 $E_{max}^{UV}=40$~eV.  But
this value is not well constrained since we don't have enough far-UV
observational points to determine overall curvature of the quasar
spectrum, and since we do not consider any spectral variability.
Alternatively, the transition between power-laws can be sharp with
$E_{max}^{UV}=10$~eV. Therefore, we adopt the value of $E_{max}^{UV}$
as a free parameter of the model.

Other parameters needed for photoionisation calculations are set in a
standard way. We use the ionisation parameter defined as:
\begin{equation}
\xi = {L \over n_{0}R^2},
\label{eq:xi}
\end{equation}
where $L$ is the quasar bolometric luminosity, $n_{0}$ is the hydrogen
number density at the cloud illuminated surface, and $R$ is the
distance of the intrinsic absorber from an UV/X-ray source, i.e.
quasar nucleus.  In all computations presented below, we calculate
grid of models for different $\xi$, $n_{0}$ and $R$, fixing the total
quasar bolometric luminosity at $L = 10^{46}$~erg~s$^{-1}$.  For
photoionisation modelling it is no important what value of $L$ we
assume, since it can be modified by moving cloud closer or farther
from the nucleus.

\subsection{Modelling with CLOUDY}
\label{sec:cloudy}

All photoionisation models computed by {\sc cloudy} assume
transmission of UV/X-ray flux through a constant density slab, which
in such case is equal to surface hydrogen density number, $n_{0}$.  We
have calculated large grid of models for different high energy
cut-offs in UV band, $E_{max}^{UV}$, ranging from 3 eV up to 42 eV,
and different ionisation parameters, $\xi$, from $10^{-2}$ up to $2.5
\times 10^{4}$.

The calculations were repeated for several distances, R, from the
quasar nucleus, equal: log~$(R)=$ 16, 16.7, 17, 18, and 20 [cm].  The
choice of distances was made to cover all expected locations of the
absorber, either close to the nucleus, at $R= 0.00324$~pc, or far away
from the central black hole, at $R= 32.4$~pc.  Since bolometric
luminosity is the same for all models, the density of the slab is
calculated from Eq.~(\ref{eq:xi}), and spans range from $10^2$ up to
$10^{14}$~\cmt.

Since column densities of ions can vary rapidly approaching the
hydrogen ionisation front, located typically at log(\NHt~)~$\sim
20.6$~[\cmd] \citep[see Fig. 1 in][]{korista2008}, we consider two
cases of clouds.  One optically thin, with total hydrogen column
density equal to log(\NHt~)~$=19$~[\cmd], and one optically thick,
with column density log(\NHt~)~$=22$~[\cmd].

In Fig.~\ref{fig:col19} we present the ratios of ionic column
densities of \CIV\ and \HI\ for different ionisation parameters and UV
high energy cut-offs, for clouds with total column density
log(\NHt~)~$=19$~[\cmd].  Uppermost panel shows results for absorber
located farthest away from a nucleus at log~$(R)=20$~[cm]. Lower
panels represent absorbers closer to the quasar centre, and distances
are marked in their upper-left corners.  The value of ionisation
parameter is shown on left vertical axis, the $E_{max}^{UV}$ on the
horizontal axis, while hydrogen density number, equal $n_0$, on right
vertical axis.  Colour contours depict the values of ratio marked in
upper-right corners of each panel.

Clouds located farther from the UV/X-ray source have low densities,
and the ratio of \NCH\ never exceeds 11.  The highest value of \NCH\
is achieved for absorber located at log~$(R)=16.7$~[cm] and reaches
18.34.

The same grid of models but for log(\NHt~)~$=22$~[\cmd] is presented
in Fig.~\ref{fig:col22}. Only for the dense cloud, $n_0=10^{12}$~\cmt,
located close to the nucleus, log~$(R)=16.7$~[cm], the ratio of \NCH
reaches 20, which agrees with our observational result presented in
Tab.~\ref{tab:ratio}, first row, so we accept this model for farther
considerations in Sec.~\ref{sec:distance} below.

\begin{figure}
\epsfxsize=8.8cm \epsfbox[50 50 380 680]{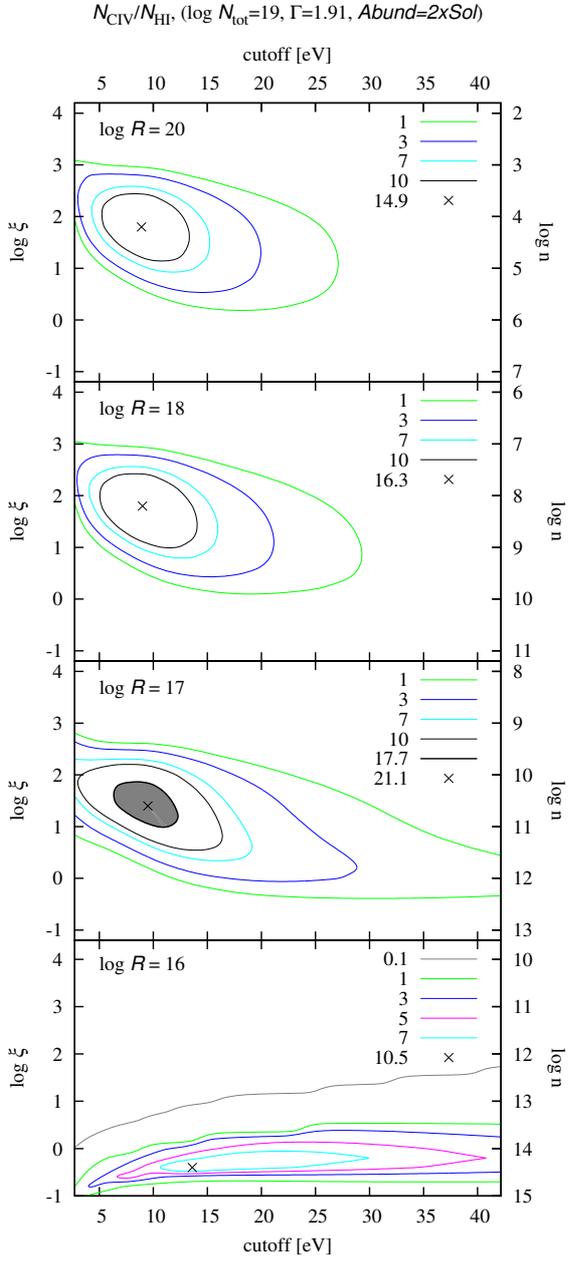}
\caption{Contour plots for the ratio of \CIV\ to \HI\ column densities
  from {\sc cloudy} for clouds with total column density
  log(\NHt~)~$=19$~[\cmd], and with larger carbon abundance by a
  factor of 2.}
\label{fig:car2}
\end{figure}

Finally, in Fig.~\ref{fig:car2}, we present the set of clouds similar
like in Fig.~\ref{fig:col19} computed with twice higher abundance of
carbon. We see, that the distance for which the ratio of \NCH\ is
above 20 appeared for log~$(R)=17$~[cm], so the distance increased by
a factor of 2. For carbon abundance five times higher than solar in
each distance there is a possibility to achieve an absorber with \NCH\
above 20, and we are not able to find one photoionisation model which
describes observations, therefore for further comparison we use only
clouds with solar metallicity.

\subsection{Modelling with TITAN}
\label{sec:titan}

One of the key issues in understanding of the nature of the absorbing
medium is its clumpiness. A clumpy medium can spontaneously form out
of a continuous wind, if the condition for thermal instability is
satisfied.  An irradiated medium may indeed be a subject of thermal
instability, as pointed out by \citet{krolik1981}, so a cold dense
cloud can coexist with a hot rare medium under constant pressure and
the size of the cloud is roughly determined by the Field length
\citep{field1965,begelman1990,torricelli1998,rozanska1999}.  It is
therefore of interest to check whether the observed systems of clouds
are consistent with the requirement of the thermal instability.

For that purpose, we consider the absorption by a cloud under constant
pressure using the code {\sc titan}, developed by \citet{dumont2000}.
The code is designed to calculate the transmission of radiation
through the cloud of considerable optical depth, since both the
continuum and lines are calculated using the full radiative transfer
instead of escape probability formalism. The density profile within
the cloud is calculated self-consistently from the requirement of the
total (i.e. radiation + gas) pressure being constant across the cloud
\citep{rozanska2006}.  Since such numerical computations are
time-consuming, we do not calculate an extended grid of models, but we
aim to reproduce the observed \CIV\ to \HI\ ratio of \HS\ .

\begin{figure}
\epsfxsize=8.8cm \epsfbox[90 360 625 700]{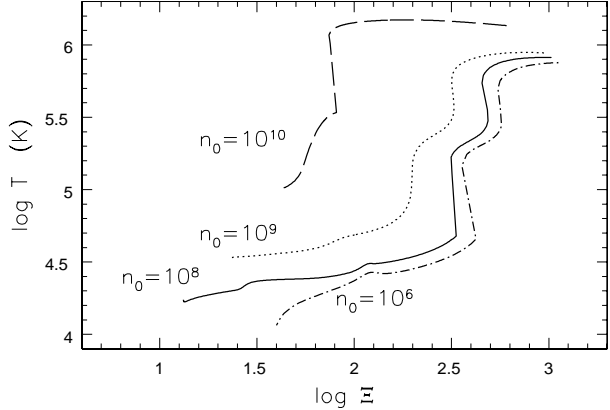}
\caption{The stability curve, i.e. temperature versus ionisation
  parameter $\Xi$ for constant pressure models computed by {\sc titan}
  code.  Different clouds are computed for the same ionisation
  parameter $\xi=10^5$ (see Eq.~(\ref{eq:xi})) at the cloud surface, and
  the same total column density \NHt~$=10^{22}$~\cmd. Different
  hydrogen number densities on the cloud surfaces are presented by:
  long-dashed line - $n_0=10^{10}$, dotted line - $n_0=10^{9}$, solid
  line - $n_0=10^{8}$, and dashed-dotted line - $n_0=10^{6}$~\cmt.}
\label{fig:TXi}
\end{figure}

We consider a range of the total column densities from $10^{21}$ up to
$10^{23}$~\cmd.  Since we compute transfer through the cloud, we are
able to make a plot of the ionisation parameter $\Xi$ introduced by
\citet{krolik1981}:
\begin{equation}
\Xi = \frac {F_{ion}}{cP_{gas}}= \frac {P_{rad}} {P_{gas}},
\label{eq:xibig} 
\end{equation}
which shows best the thermal instability region. For each of the model
computed by {\sc titan} we fix cut-off energy in UV/EUV band on the
value $E_{max}^{UV}=40$~eV.  The free parameters of our model are the
ionisation parameter $\xi$, and density number $n_0$, both at the
cloud surface, and the total column density \NHt.  The method is
illustrated in Fig.~\ref{fig:TXi} where we show example of the
temperature $T$ vs. $\Xi$ relation for several models.  For the
purpose of this paper we have calculated several clouds with following
sets of parameters: cloud I: $n_0=10^{10}$\cmt\ , $\xi=10^6$ ,
\NHt~$=10^{23.3}$~\cmd ; cloud II: $n_0=10^{10}$~\cmt\ , $\xi=10^4$ ,
\NHt~$=10^{22.7}$~\cmd ; cloud III: $n_0=10^{8}$~\cmt\ , $\xi=10^6$ ,
\NHt~$=10^{22}$~\cmd ; cloud IV: $n_0=10^{9}$~\cmt\ , $\xi=10^4$ ,
\NHt~$=10^{22}$~\cmd ; and cloud V: $n_0=10^{10}$~\cmt\ , $\xi=5
\times 10^3$, \NHt~$=10^{22}$~\cmd .

Even for the soft spectrum, in case of HS 1603+382, the instability is
still present, contrarily to the previous expectations
\citep{krolik1981}, but it appears at much lower optical depth since
the Compton temperature of the incident radiation flux is considerably
lower. Such a lower optical depth is in agreement with relatively low
column densities of specific ions in comparison with those detected in
a number of Seyfert 1 galaxies \citep[][and references
therein]{rozanska2006}.

\begin{figure}
\epsfxsize=8.8cm \epsfbox[90 490 625 700]{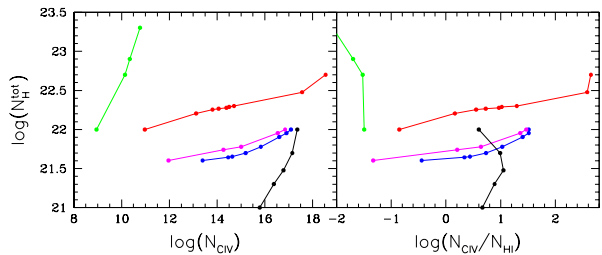}
\caption{Total column density versus column density of \CIV\ - left
  panel, ratio of \NCH\ - right panel, for different constant pressure
  clouds. Green solid line represents models for $n_0=10^{10}$\cmt\
  and $\xi=10^6$, red line represents models for $n_0=10^{10}$~\cmt\
  and $\xi=10^4$, magenta line for $n_0=10^{8}$~\cmt\ and $\xi=10^6$,
  blue line for $n_0=10^{9}$~\cmt\ and $\xi=10^4$, and black line for
  $n_0=10^{10}$~\cmt\ and $\xi=5 \times 10^3$. Each computed cloud is
  represented by point on those panels.}
\label{fig:kontury}
\end{figure}

In Fig.~\ref{fig:kontury}, we present total column density of the
cloud versus column density of \CIV\ in the left hand panel, and the
ratio of \NCH\ in the right panel for several models.  The column
density of \CIV\ is the monotonic function of total column density,
but ratio tends to have maximum.  The best representation of the
observed \CIV\ content was obtained for $n_0=10^{10}$, $\xi=10^4$, and
\NHt~$=2 \times 10^{22}$~\cmd\ - marked as a red point in
Fig.~\ref{fig:kontury}.  For this model we achieved
log(\NC~)~$=14.71$~[\cmd] and \NCH~$=19.70$, which is very similar to
the values observed in system A1.

Note, that for constant pressure clouds, there are models presented by
magenta and blue points with, $n_0=10^{8}$ and $10^9$~\cmt\
respectively, with ratio of \NCH\ approaching 20, but at the same time
column density of \CIV\ is huge i.e. log(\NC~)~$=16-17$~[\cmd] for
those models, and exceeds the value appropriate for the A1.

%______________________________________________________
\section{Distance to the A1 absorber}
\label{sec:distance}

In the aim to estimate the distance to the A1, we have to know
bolometric luminosity of the quasar.  Then, the photoionisation modelling
presented above together with the observational constrains allows us to
conclude that the best representation of system A1 is cloud with $
\xi= 7$ and $n_0=10^{12}$ from {\sc cloudy} and with $\xi=10^4$ and
$n_0=10^{10}$ from {\sc titan}.  The huge difference in ionisation
parameter on the cloud surface is because a cloud computed by {\sc
  titan} is stratified and contains both hot and cold phases (see
Fig.~\ref{fig:TXi}), and a cloud computed by {\sc cloudy} contains
only cold phase of the absorber.  Finally, after manipulating of
equation (\ref{eq:xi}) we derive the distance.

The most important is thus to estimate bolometric luminosity of the
considered quasar.  \citet{misawa2005} gave the value of bolometric
luminosity based on the prescription of \citet{narayanan2004}, where:

\begin{equation}
L_{bol} \approx 4.4 ~ \lambda ~ L_{\lambda} ~~~~~~~~~  at ~~~~ \lambda=1450 \AA\  .
\label{eq:bol}
\end{equation}

For \HS\, \citet{misawa2005} have obtained $L_{bol} = 2.5 \times
10^{48}$ ergs s$^{-1}$.  We did similar evaluation, besides that we
have adopted the value of $f_{\nu}^{obs}=550$ $\mu$Jy at
$\lambda=1450$~\AA, as reported for this quasar by \citet{scott2000}
(see their Tab.~3).  Assuming standard cosmological parameters:
$H_0=72$ \kms Mpc$^{-1}$, $\Omega_M=0.3$, and $\Omega_{\Lambda}=0.7$
for luminosity distance, and using Eq.~(\ref{eq:bol}), we have computed
that $L_{bol}=7.7 \times 10^{47}$ erg s$^{-1}$.  This value is lower
from that obtained by \citet{misawa2005}, by a factor of 3.25. Note,
that such derivation of bolometric luminosity has the factor of two
uncertainty.

\begin{table}
\begin{center}
  \caption{Distance derivations based on {\sc cloudy} and {\sc titan}
    modelling for the A1 in \HS, for solar metallicity.  Middle rows
    contain results based on bolometric luminosity calculated in this
    paper from Eq.~(\ref{eq:bol}) using flux reported by
    \citet{scott2000}. Bottom rows - the one calculated by
    \citet{misawa2005}. Ionisation parameter, $\xi$ on the surface of
    the cloud drops significantly inside the cloud for the {\sc titan}
    model, and remains constant for {\sc cloudy} model. See text for
    explanation.}
\label{tab:dist1}
\begin{tabular}{llll} % No vertical elements in tables.
\hline \hline 
 Par.   & Unit   & CLOUDY & TITAN \\
\hline
 $\xi $ & [erg s$^{-1}$ cm$^{-1}$]   &  7 & $10^4$ \\  
$n_0$ & [\cmt]   & $10^{12}$ & $10^{10}$ \\
$E^{UV}_{max}$ &[eV]  & 10 & 40 \\
log(\NC~)  & [\cmd] &  14.97 & 14.71\\
\NCH\ & & 20.37 & 19.70 \\
\hline  
\hline 
   $ L_{bol} =7.7$ & $ \times 10^{47}$ [erg s$^{-1}$]  & This paper &  \\
\hline 
log~$(R)$ & [cm]        & 17.52 & 16.94\\
 R & [pc]              & 0.106 & 0.028\\
\hline
\hline  
   $ L_{bol} = 2.5$  & $\times 10^{48}$ [erg s$^{-1}$]  & Misawa 05 & \\
\hline 
log~$(R)$ & [cm]        & 17.78  & 17.20 \\
 R & [pc]              & 0.192 & 0.051 \\
\hline \hline
\end{tabular}
\end{center}
\end{table}

Taking into account both values of bolometric luminosities, we derive
the distance to the A1 for models computed by {\sc cloudy} and {\sc
  titan}.

Regardless of the value of $L_{bol}$, in each case the A1 absorber is
located very close to the quasar nucleus within $R=0.1-0.2$ pc.  Our
results differ from these obtained by \citet{misawa2005,misawa2007}
using variability method \citet{hamann97}. Assuming that variability
is caused by the change of ionisation, they found for $\Delta t \sim
0.36$ yr the location of system A within $r<6$ kpc, and for 1.2 yr the
location of the mini-BAL in system A within $r<8$ kpc
\citep{misawa2005}.

Nevertheless, our results are in agreement with dynamical model of
\citet{murray95} adopted for \HS\ by \citet{misawa2005} (see Sec.
6.2). Using the relation for the radius of the gas parcel relative to
its launch radius, they have obtained constraint of $r<0.2$ pc for
system A.
 All distance derivations are given in Table~\ref{tab:dist1}.

Our results are less restrictive if the lower value of the \CIV\ to
\HI\ ratio is used, or metallicity is allowed to be higher.  The
increase in the metallicity by a factor of two moves the upper limit
for the distance also by a factor of 2.  However, if we adopt lower
cloud covering factor and, in consequence lower value of the \NCH\
ratio, the upper limit for the cloud distance moves to the 30 pc.  The
lower limit to the distance remains practically unchanged.

\section{Connection with an accretion disk and BLR}
\label{sec:blr}

The distance to the A1 absorber of \HS\ derived by us suggests that
the absorbing material may be connected with the wind from an
accretion disk atmosphere. Upper atmospheric layers can be quite dense
up to $n_H \sim 10^{14}$~\cmt, depending on the distance from the
central black hole as pointed by \citet{hrynio2011}. Such density can
be calculated by solving an accretion disk vertical structure
parametrised by the black hole mass, its spin and the accretion rate
\citep{rozanska99b}.  We can properly derive density at $\tau \sim
2/3$, since we assume diffusion approximation for radiation when
solving the disk vertical structure.  This approximation is valid for
the disk interior, but does not solve properly the structure of outer
atmosphere at $\tau \ll 1$.  However, the density at $\tau=2/3 $ is
representative as an initial density in the disk wind, which supplies
matter to the BLR or other intrinsic absorber.

\begin{figure}
\epsfxsize=8.8cm \epsfbox[35 0 590 410]{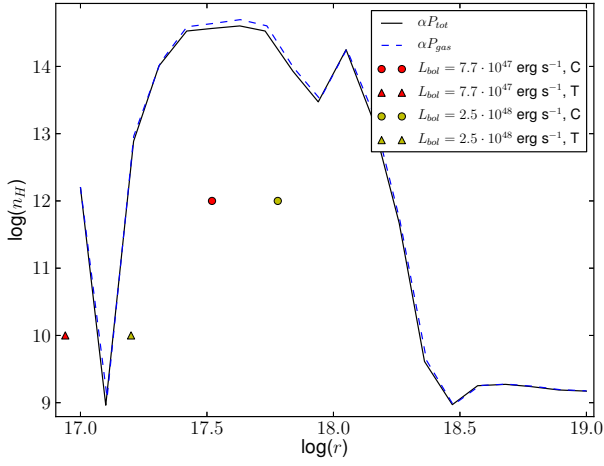}
\caption{The value of number density on the disk surface, calculated
  from an accretion disk vertical structure (see
  Sec.~\ref{sec:distance} for explanation), versus the distance from
  the quasar centre. Two radial profiles (blue and black) are
  calculated for two viscous heating prescription marked in the left
  upper corner.  Points represent values achieved for both parameters
  from photoionisation modelling.  Clouds computed with {\sc cloudy
    (c)}, are marked by circles, and those computed with {\sc titan
    (t)} - by triangles.  Red points represent clouds with bolometric
  luminosity calculated by us, while green points - by
  \citet{misawa2005}.}
\label{fig:vert}
\end{figure}

Radial profile of the density atmosphere at the optical depth $\tau
\sim 2/3$, in an accretion disk around non-rotating black hole in the
centre of weak line quasar SDSS J0945533.99+100950 is presented in
\citet{hrynio2011}. Authors have shown visible bump in disk's density
distribution caused by the drop of the opacity in the region where the
temperature is too low for significant contribution of electron
scattering and free-free processes \citep{hure1994} but still too high
for opacity due to the dust, so only molecules are the source of
opacity.  Sudden drop of the volume hydrogen density from $10^{14}$ to
$10^9$~\cmt\ occurs at log~$(R) \sim 18$~[cm] (see Fig.~5 in
\citet{hrynio2011}), and coincides perfectly with the distance to BLR
for this object.

We have made the same calculations for the mass of the black hole
appropriate for \HS\ , derived from bolometric luminosity $L_{bol} =
7.7 \times 10^{47}$ erg s$^{-1} $, assuming mass accretion rate equal
Eddington, black hole mass is $ M_{BH} =5.26 \times 10^9 M_{\odot}$.
Furthermore, we have estimated black hole mass in this quasar using
\CIV\ line \citep{vester2006}, which gives similar value $ M_{BH} =5.4
\times 10^9 M_{\odot}$, with uncertainty of one order of magnitude.

Taking into account bolometric luminosity derived by us, we consider
an accretion disk around black hole of the mass $5.26 \times 10^9
M_{\odot}$, and we assume Eddington accretion rate for the gas.  For
such disk we have computed radial profile of the density $n_H$ at $
\tau=2/3 $, assuming two different viscous heating prescriptions.
First, we assume that viscous heating is proportional only to the gas
pressure, so we do not expect any radiation pressure instabilities.
Second, we assume that viscous stress is proportional to the total
pressure.

Fig.~\ref{fig:vert} presents radial profiles of $n_H$, for two
different viscosity prescriptions. The drop in the density of the disk
atmosphere at the radius log$(R) \sim 17$~[cm] is related to the
partial ionisation of the hydrogen in the disk atmosphere and a very
strong inversion in the vertical density profile due to radiation
pressure. This partial ionisation effect is not related to the overall
disk instability since the ionisation instability operates when there
is partial ionisation of the disk interior (i.e. further out).  What
is more, if we change the viscosity prescription from $\alpha P_{tot}$
to $\alpha P_{gas}$, the plot practically does not change, although
the integrated disk surface density, $\Sigma$, changes more than by an
order of magnitude.  Such a strong density inversion in the disk
atmosphere can lead to local instabilities and an outflow.
Additionally, all points from our modelling are marked on this diagram.
They are in good agreement with values predicted for outer layers of
an accretion disk atmosphere around black hole with mass estimated by
us from observational constrains.

Furthermore, following equation~(3) in \citet{pian2005}, we can
calculate the distance to the BLR in \HS, from the continuum
luminosity at 1350 \AA.  Estimation is rather crude, since absorption
can be present at this wavelength, but we have got the value of
log$(R_{BLR})=18.75$~[cm].  Together with our distance derivation to
the A1 absorber, we conclude that intrinsic absorption is located in
radial direction closer to the nucleus than a possible BLR in \HS.

%_______________________________
\section{Conclusions}
\label{sec:summary}

We have presented photoionisation modelling of intrinsic absorber in
quasar \HS. Using observations from optical/UV/X-ray energy bands, we
have constructed the shape of continuum, which interacts with
absorbers on the way from the nucleus.  Double power-law with
exponential cut-offs and related observed optical to X-ray slope
$\alpha_{ox} = 1.49$, was adopted as an input to photoionisation
modelling of the first cloud in system A.
 
For several lines of system A1, reported in Paper I, we have found
very high ratio of column densities of \CIV\ to \HI\ .  The value of
this ratio in some cases exceeds 20.  Our photoionisation computations
done by {\sc cloudy} and {\sc titan} show that this value, together
with intrinsic SED of \HS, allowed us to choose a single solution
which explains all observational constrains.

From modelling done by {\sc titan}, we found that total column density
of absorbing system is $2 \times 10^{22}$~\cmd.  This value is quite
close to the maximum column density in a constant pressure medium due
to the thermal instability. It is one order of magnitude lower that
the values derived for Seyfert galaxies \citep{rozanska2008}, since
the intrinsic spectrum of \HS\ is much softer, with a very weak X-ray
tail. This coincidence supports the view that intrinsic absorbers do
form in the wind as a result of thermal instability and the constant
pressure model is better suited to interpret the data than the
constant density one.

The hydrogen number density derived from modelling is rather high, of
the order of $n = 10^{10}, 10^{12}$~\cmt, which is consistent with
previous density estimations of UV intrinsic absorbers in NLS1
galaxies (see \citet{karen2004}). On the other hand, our result is
inconsistent with number density found by \citet{misawa2005} using
variability method.

Such high value of number density gives the location of absorbing gas
quite close to the nucleus within $R=0.1-0.2$ pc. This location
supports scenario that UV/X-ray outflows are in the form of winds from
accretion disk atmospheres with similar number densities. Furthermore,
we have shown that the A1 absorber is located closer to the nucleus
than the BLR for a given bolometric luminosity of \HS.

The results, however, are very sensitive to the measured \CIV\ to \HI\
ratio, which is sensitive to the determination of the covering factor,
very difficult for hydrogen line. For the same covering factor for
\HI\ and \CIV\ of 0.36 in the Keck 2001 data, the ratio is much lower
than measured in Paper I and the upper limit for the distance is moves
to 30 pc. Also assumption of higher metallicity increases the upper
limit for the absorber distance.

Close location of the absorber system is in contradiction with
\citet{misawa2005,misawa2007} who found that system A is located
within $r=6$ kpc, but it agrees with dynamical model of
\citep{murray95} adopted for \HS\ by \citep{misawa2005} (see Sec.
6.2). Using the relation for the radius of the gas parcel relative to
its launch radius, they have obtained constraint of $r<0.2$ pc for
system A.

The outflow velocities of system A are rather high, up to few
thousands \kms, which is consistent with the winds detected in X-rays
in several NLS1. Therefore, HS1603+3820 can be considered as a high
redshift analog of NLS1 galaxies.

%
%_____________________________________________________________
\section*{Acknowledgements}

We thank A. Gawryszczak for his help with model computations.  Support
for this work was provided by grant 2011/03/B/ST9/03281,
2012/04/M/ST9/00780, and NN203 581240 the Polish
National Science Center. Some of the data presented
in this paper were obtained from the Multimission Archive at the Space
Telescope Science Institute (MAST). STScI is operated by the
Association of Universities for Research in Astronomy, Inc., under
NASA contract NAS5-26555. Support for MAST for non-HST data is
provided by the NASA Office of Space Science via grant NAG5-7584 and
by other grants and contracts.


\bibliographystyle{elsarticle-harv}
\bibliography{RozanskaNewA_refs}


\end{document}